\documentstyle[12pt]{article}
\begin{document}
\large
\begin{center}
{\bf Oscillations of $K^0$ Mesons in the Phase Volume Approach in the
Standard Model and in the Model of Dynamical Analogy of
the Cabibbo-Kobayashi-Maskawa Matrices}\\

Kh.M. Beshtoev\\

Joint Instit. for Nucl. Research., Joliot Curie 6, 141980 Dubna,
Moscow region, Russia;
Scientific Research Institute of Applied
Mathematics and Automation of the Kabardino-Balkarian Scientific Center of
RAC, Shortanova 89a, 360017 Nalchik, Russia\\
\end{center}
\large
{\bf Abstract}
\par
The elements of the theory of dynamical expansion of the theory of weak
interaction working at the tree level, i.e. the model of dynamical analogy
of Cabibbo-Kobayashi- Maskawa matrices, are given.
\par
The equation for mass difference of $K^0_1, K^0_2$ mesons or the length
of $K^0, \bar K^0$ meson oscillations in the phase volume approach in
the frame work of the standard model of the weak interactions and
the model of dynamical analogy of Cabibbo-Kobayashi-Maskawa matrices
is calculated (in this model the oscillations of $K^0, \bar K^0$
mesons arise at violation of strangeness by $B$ bosons).
Comparison of the length of oscillations in the diagram and phase volume
approaches was done. The length of $K^0  \leftrightarrow \bar K^0$
oscillations in the phase volume approach  is much more
than the length of these oscillations in the diagram approach.\\
\par
\noindent
PACS: 12.15 Ff Quark and lepton masses and mixing.\\
PACS: 12.15 Ji Aplication of electroweak model to specific processes.\\

\section{Introduction}
\par
At the present time the theory of electroweak interactions
has a status of theory which is confirmed with  a  high  degree  of
precision. However, some experimental  results  (the  existence  of the
quarks and the leptons families, etc.)
have not got any explanation in the framework of
the theory.  One part of the electroweak theory is the  existence of
quark  mixings introduced   by   the Cabibbo-Kobayashi-Maskawa
matrices (i.e., these  matrices  are  used for parametrization of the quark
mixing).
\par
In previous works [1] a dynamical mechanism of quark mixing
by the  use  of four  doublets  of massive  vector carriers  of weak
interaction $B^{\pm}, C^{\pm}, D^{\pm}, E^{\pm} $,
i.e. expansion of the standard theory of weak interaction (the theory
of dynamical analogy of the Cabibbo-Kobayashi-Maskawa matrices) working
at the tree level, was proposed.
\par
The vacuum oscillation of neutral $K$ mesons is well investigated at
the present
time [2]. This oscillation is the result of $d, s$ quark mixings and
is described by Cabibbo-Kobayashi-Maskawa matrices [3, 4]. The angle
mixing $\theta$ of neutral $K$ mesons is $\theta = 45^O$ since $K^o,
\bar K^o$
masses are equal (see $CPT$ theorem). Besides, since their masses are equal,
these oscillations are real, i.e. their transitions to each other are going
without suppression.
\par
This work is devoted to the study of $K^0, \bar K^0$ oscillations in
the phase volume approach in the frame work of the standard model and the
model of dynamical analogy of the Cabibbo-Kobayashi-Maskawa matrices.
\par
At first, we will give the general elements of the model
of dynamical analogy of the Cabibbo-Kobayashi-Maskawa matrices,
the $K^0, \bar K^0$ oscillations will be considered in the frame work of
these models and then the comparison of length of oscillations
in the diagram and phase volume approaches is fulfilled.

\section {The Theory of Dynamical Analogy of the Cabibbo-Kobayashi-Maskawa
matrices}

In the case of three  families of quarks the current $J ^{\mu }$
has  the following form:
$$ J^{\mu } = ( \bar{u}  \bar{c}  \bar{t} )_{L} \gamma ^{\mu} V \left
( \begin{array}{c}
d\\ s\\ b\\ \end{array} \right)_{L}
\eqno(1)
$$
\begin{displaymath} {V =
\left( \begin{array}{ccc} V_{ud}& V_{us}& V_{ub}\\ V_{cd}& V_{cs}& V_{cb}\\
V_{td}& V_{ts}& V_{tb} \end{array} \right)}  .
\end{displaymath}
where $V$ is a Kobayashi-Maskawa  matrix [4].
\par
Mixings of the $d, s, b$ quarks
are not related to the weak interaction (i.e., with
$W^\pm, Z^o$ bosons exchanges). From equation (1) it is well seen that
mixings of the $d, s,b$ quarks and exchange of $W^\pm, Z^o$- bosons take place
in an independent manner (i.e., if matrix $V$ were diagonal, mixings of
the $d, s, b$ quarks would not have taken place).
\par
If the mechanism of this mixings is realized independently of the weak
interaction ($W^\pm, Z^o$-boson exchange) with a probability determined by
the mixing angles $\theta, \beta, \gamma, \delta$ (see below), then this
violation could be found in the strong and electromagnetic interactions of
the quarks as clear violations of the isospin, strangeness and beauty.
But, the available experimental results show that there are no
clear violations of the number conservations in strong and electromagnetic
interactions of the quarks. Then we must relate the non-conservation of the
isospins, strangeness and beauty (or mixings of the $d, s, b$ quarks) with
some type of interaction mixings of the quarks. We can do it introducing
(together with the $W^\pm, Z^o$-bosons) the heavier vector bosons
$B^\pm, C^\pm, D^\pm, E^\pm$ which interact with the $d, s, b$ quarks with
violation of isospin, strangeness and beauty.
\par
We shall choose  parametrization of  matrix $V$  in
the form offered by Maiani [5]
\par
\begin{displaymath}{V = \left( \begin{array} {ccc}1& 0 & 0 \\
0 & c_{\gamma} & s_{\gamma} \\ 0 & -s_{\gamma} & c_{\gamma} \\
\end{array} \right) \left( \begin{array}{ccc} c_{\beta} & 0 &
s_{\beta} \exp(-i\delta) \\ 0 & 1 & 0 \\ -s_{\beta} \exp(i\delta) &
0 & c_{\beta} \end{array} \right) \left( \begin{array}{ccc} c_{\theta}
& s_{\theta} & 0 \\ -s_{\theta} & c_{\theta} & 0 \\ 0 & 0 & 1
\end{array}\right)} , \end{displaymath}
\par
$$
c_{\theta} = \cos {\theta } , s_{\theta} =\sin{\theta} , \exp(i\delta)
= \cos{\delta } + i \sin{\delta} .
\eqno(2)
$$
\noindent
To the nondiagonal terms in (2), which are  responsible for  mixing  of
the $d,s,b$- quarks and $CP$-violation in the three matrices,
we shall  correspond four doublets of vector bosons
$B^{\pm},C^{\pm},D^{\pm},E^{\pm} $ whose contributions
are parametrized  by four  angles $\theta ,\beta ,\gamma ,\delta $ .
It is supposed that the  real
part of $Re(s_{\beta} \exp(i\delta)) = s_\beta \cos{\delta} $
corresponds to the vector boson $C^{\pm}$ , and  the imaginary part
of $Im(s_{\beta } \exp(i\delta)) = s_{\beta}\sin \delta$ corresponds
to the vector boson $E^{\pm}$ (the couple constant  of $E$  is
an imaginary value !). Then, when $q^{2}<< m^{2}_{W}$ , we get:
\par
$$
\tan{\theta } \cong \frac{m^{2}_{W} g^{2}_{B}}{ m^{2}_{B} g^{2}_{W}},
\qquad
\tan{\beta} \cong \frac{m^{2}_{W} g^{2}_{C}}{ m^{2}_{C} g^{2}_{W}} ,
$$
$$
\tan{\gamma} \cong \frac{m^{2}_{W} g^{2}_{D}}{ m^{2}_{D} g^{2}_{W}} ,
\qquad
\tan{\delta} \cong \frac{m^{2}_{W} g^{2}_{E}}{m^{2}_{E} g^{2}_{W}} .
\eqno(3)
$$
\par
\noindent
If $g_{B^{\pm}} \cong  g_{C^{\pm}} \cong g_{D^{\pm}} \cong  g_{E ^{\pm}}
\cong g_{W^{\pm}}$ , then
$$ \tan{\theta } \cong \frac{ m^{2}_{W}}{ m^{2}_{B}} ,
\qquad
\tan{\beta} \cong \frac{m^{2}_{W}}{ m^{2}_{C}}  , $$
$$\tan{\gamma} \cong \frac{m^{2}_{W}}{ m^{2}_{D}}  ,
\qquad
\tan{\delta} \cong \frac{m^{2}_{W}}{ m^{2}_{E}}  .
\eqno(4)
$$
\par
Concerning the  neutral  vector  bosons $B^{0} , C^{0} , D^{0}, E^{0}$, the
neutral scalar bosons $B^{'0},C^{'0},D^{'0} ,E^{'0}$and the GIM
mechanism  [6], we can repeat the same
arguments which were given in the previous work [1].
\par
Using the data from [2] and equation (4), we have obtained the following
masses for $B^\pm, C^\pm, D^\pm, E^\pm$-bosons:
\par
$
m_{B^{\pm}} \cong  169.5 \div 171.8\hbox{ GeV.};\\
$
\par
$
m_{C^{\pm}} \cong  345.2 \div 448.4\hbox{ GeV.};\hspace{8cm}(5)\\
$
\par
$
m_{D^{\pm}} \cong 958.8 \div 1794$ GeV.;\\
\par
$
m_{E^{\pm}} \cong  4170 \div 4230\hbox{ GeV..}\\
$

\section{Oscillations of Neutral $K^0, \bar K^o$ Mesons in the
Phase Volume Approach in the Standard Model and in the Model
of Dynamical Analogy of Kabibbo-Kobayashi-Maskawa Matrices}

The oscillations of $K^0, \bar K^0$ mesons are characterized by two
parameters: angle mixing--$\theta$ and length-- $R$ or time
oscillations--$\tau$.
\par
At first, we will consider mixings of $K^0$ mesons in the phase volume
approach and
obtain the expression for matrix elements and the time of
$K^0, \bar K^0$ transitions arising for existence of the strangeness
violation
of the weak interaction through  $sin\theta W$ or $B$ exchange in the
standard or in our model of dynamical analogy of Kabibbo-Kobayashi-Maskawa
matrices. The general scheme of $K^0, \bar K^0$ meson oscillations is given
in the end of the article.
\par
In further considerations, for transition from the standard model to our
model, the following values are used for $sin\theta, cos\theta$ and $G_F$:
$$
sin\theta \cong \frac{m^2_W g^2_B}{m^2_B g^2_W} \cong
\frac{m^2_W }{m^2_B} ,
$$
$$
cos^2 \theta = 1 - sin^2 \theta ,
\eqno(6)
$$
$$
G^2_F = \frac{g^{2}_{W}}{32 m^{2}_{W}} .
$$

\par
{\bf A) Mixings of $K^0, \bar K^0$ Mesons} \\

\par
Let us consider mixings of $K^0, \bar K^0$ mesons.
$K^0$ and $\bar K^0 $ consist of $\bar d, s, d, \bar s,$ quarks and
have the same masses (it is a consequence of the $CPT$
invariance) but their strangenesses are different $s_{K^0} = -1,
s_{\bar K^0}=1$. Since the weak interaction through $B$ boson exchanges
changes the strangeness, then the $K^0, \bar K^0$ are mixed.
\par
The mixings of $K^0$ mesons can be considered by using the
following nondiagonal
mass matrix of $K^0$ mesons:
$$
\left(\begin{array}{cc}  m^2_{K^0} & m^2_{K^0 \bar K^0}\\
m^2_{\bar K^0 K^0} & m^2_{\bar K^0} \end{array} \right) .
\eqno(7)
$$
Since $m^2_{K^0} = m^2_{\bar K^0}$, the angle $\theta'$ of $K^0, \bar K^0$
mixing, or the angle rotation for diagonalization of this matrix

\begin{displaymath} \left(\begin{array}{cc}  m^2_{1} & 0 \\
0 & m^2_{2} \end{array} \right) , \end{displaymath}
is given by the expression:
$$
tg2\theta' = \frac{m^2_{K^0 \bar K^0}}{m^2_{K^0} - m^2_{\bar K^0}},
$$
and is equal to ($\theta' = )  \frac{\pi}{4}$ and
$$
m^2_{1,2} = \frac{1}{2}((m^2_{K^0} - m^2_{\bar K^0}) \pm
\sqrt{(m^2_{K^0} - m^2_{\bar K^0})^2 + 4(m^2_{K^0 \bar K^0})^2}),
\eqno(8)
$$
$$
m^2_1 - m^2_1 = m^2_{K^0 \bar K^0} .
\eqno(9)
$$
Then the following new states $K^0_1, K^0_2$ arise:
$$
K^0_1 = \frac{K^0 + \bar K^0}{\sqrt{2}} , \hspace{.5cm} K^0_2 =
\frac{K^0 - \bar K^0}{\sqrt{2}}
\eqno(10)
$$
\par
In the quark model we can take a mass matrix with masses in the first degree
($m^2_{a b} \rightarrow m_{a b}$).
\par
Below the $CP$ --invariance is supposed to be strong
conserved (the consequences which arise of the $CP$ violation were
discussed in work [7] ),
and then the following decays are possible:
($CP$ parity  of $K^0_1$ is $P(K^0_1) = +1$, and $\bar K^0_2$ is $P(\bar
K^0_2) = -1$)
$$
K^0_1 \to 2\pi, \hspace{.5cm} K^0_2 \to 3\pi .
\eqno(11) \\
$$

{\bf B). Estimation of Probability or Time
of  $K^{0} \leftrightarrow \bar K^{0}$
Meson Transitions in the Phase Volume Approach}\\

\par
The calculation of probability or time of  $K^{0}
\leftrightarrow \bar K^{0}$ meson transitions will be executed in
the framework of the weak quark interactions by using the following
Feynman diagram:
\vspace{1cm}

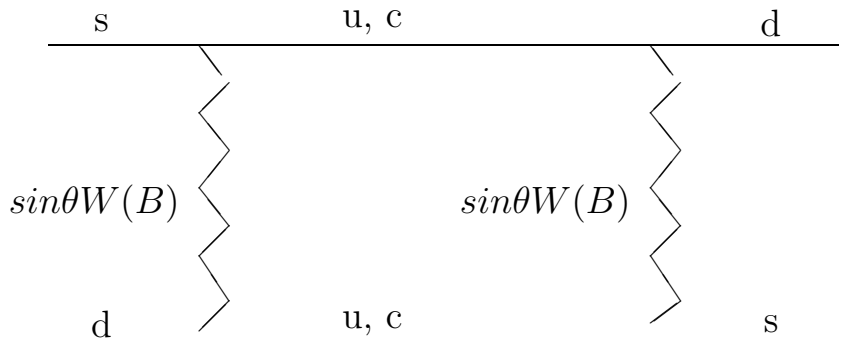
\unitlength=1.00mm
\special{em:linewidth 0.4pt}
\linethickness{0.4pt}
\begin{picture}(124.00,70.00)
\put(24.00,67.00){\line(1,0){100.00}}
\put(24.00,27.00){\line(1,0){100.00}}
\put(39.00,67.00){\line(3,-4){3.00}}
\put(39.00,29.00){\line(3,-2){3.00}}
\put(99.00,67.00){\line(3,-4){3.00}}
\put(99.00,30.00){\line(1,-1){3.00}}
\put(19.00,67.00){\line(1,0){6.00}}
\put(19.00,67.00){\line(1,0){5.00}}
\put(19.00,67.00){\line(1,0){10.00}}
\put(29.00,67.00){\line(-1,0){10.00}}
\put(25.00,46.00){\makebox(0,0)[cc]{$sin \theta W (B)$}}
\put(85.00,46.00){\makebox(0,0)[cc]{$sin \theta W (B)$}}
\put(24.00,27.00){\line(-1,0){5.00}}
\put(26.00,70.00){\makebox(0,0)[cc]{s}}
\put(115.00,70.00){\makebox(0,0)[cc]{d}}
\put(115.00,30.00){\makebox(0,0)[cc]{s}}
\put(26.00,30.00){\makebox(0,0)[cc]{d}}
\put(62.00,30.00){\makebox(0,0)[cc]{u, c}}
\put(62.00,70.00){\makebox(0,0)[cc]{u, c}}
\put(42.00,63.00){\line(1,-1){1.00}}
\put(43.00,62.00){\line(-1,-1){4.00}}
\put(39.00,58.00){\line(4,-5){4.00}}
\put(43.00,53.00){\line(-4,-5){4.00}}
\put(39.00,48.00){\line(4,-5){4.00}}
\put(43.00,43.00){\line(-1,-1){4.00}}
\put(39.00,39.00){\line(2,-3){4.00}}
\put(43.00,33.00){\line(-1,-1){4.00}}
\put(102.00,63.00){\line(1,-1){1.00}}
\put(103.00,62.00){\line(-1,-1){4.00}}
\put(99.00,58.00){\line(4,-5){4.00}}
\put(103.00,53.00){\line(-4,-5){4.00}}
\put(99.00,48.00){\line(4,-5){4.00}}
\put(103.00,43.00){\line(-1,-1){4.00}}
\put(99.00,39.00){\line(2,-3){4.00}}
\put(103.00,33.00){\line(-4,-3){4.00}}
\end{picture}

where $sin \theta W$ and $B$ are a bosons changing the strangeness;
$u, c$ are quarks (for simplification the $t$ quark is not taken into
account).
\par
We do not give here the details of the calculation on this diagram, since
they were widely
discussed in literature [8, 9]. Using the standard Feynman rules
for the weak interaction (after integrating on the inside
lines, twice using the Firtz rules and without
the outside momenta), we obtain the following expression for the amplitude
of $K^0 \to \bar K^0$ transition:
$$
M(K^0 \to \bar K^0) = -\frac{G^2_F m^2_c sin^2 \theta cos^2 \theta}{8 \pi^2}
\bar d Q_\alpha s \bar d Q^\alpha s =
\eqno(12)
$$
$$
= G \bar d Q_\alpha s \bar d Q^\alpha s ,
$$
where
$$
G = \frac{G^2_F m^2_c sin^2 \theta cos^2 \theta}{8 \pi^2}  , \qquad
Q_\alpha = \gamma_\alpha (1 - \gamma_5)  .
$$
\par
Then we use the following phenomenological expression:
$$
< 0 \mid \bar s Q_\alpha d \mid \ K^0 > = \varphi_K f_K p_\alpha ,
\eqno(13)
$$
where  $\phi_K$ is a $K$-meson wave function and $\mid
\phi_K \mid_2 = 1 $, $f_K$ is a $K$ constant decay and $f_K \cong 1.27 f_\pi$
($f_\pi \cong 125$ MeV), $p_\alpha$ is a $\pi$ four-momentum.
\par
Also, we suppose that the beginning state $\bar s Q_\alpha d$ corresponds
to $K^0$ meson.
\par
Introducing the values
$F_\alpha = f_K \phi_K p_\alpha$, $Q^\alpha = \bar d_L \gamma^\alpha u_L$,
we rewrite exp.(12) in the following form:
$$
M = G  F_\alpha Q^\alpha =  ,
\eqno(14)
$$
\par
If we use the expression $\hat p_\pi = \hat p_s + \hat p_{\bar d}$ and
the Dirac equation
$$
(\hat p - m) u = 0
$$
(where $u$ is a quark wave function), then one can rewrite Eq.(14) in the
following form:
$$
M = G f_K \phi_K (m_s + m_{\bar d}) \bar d_L
\gamma_5 s_L .
\eqno(15)
$$
Using the standard procedure for $\bar {{\mid M \mid}^2}$, one obtains
the following expression:
$$
\bar {\mid M \mid^2} = G^2 f^2_K (m_s + m_{\bar d})^2
4(p_{\bar d} p_s) \cong
$$
$$
\cong 4 G^2  f^2_K (m_s + m_{\bar d})^2 m_K E_d ,
\eqno(16)
$$
where $E_d$ is the energy of $d$ quark in $K^0$ meson ($(p_d, p_s) =
(p_d,(p_K - p_d)) = m_K E_d - m_d^2 \simeq m_d m_K,  m_d^2 \simeq 0$).
$$
E_d = m_K (1 - \frac{m_s^2}{m_K^2})^2 .
\eqno(17)
$$
\par
Then the probability $W(...)$ of $K^0 \leftrightarrow \bar K^0$ meson
transitions is
$$
W(K^0 \leftrightarrow \bar K^0) = 2 \frac{\bar {\mid M \mid^2}}{2
m_{K^0}}
\int \frac{d^3 p_d}{2 E_d (2 \pi)^3} \frac{d^3 p_s}{2 E_s (2 \pi)^3}
(2 \pi)^4 \delta( p_s + p_{\bar d} - p_\pi) =
$$
$$
= 2 \frac{\bar {\mid M \mid^2}}{4 \pi m_{K^0}}
\int \delta(E_{\bar d} + E_s - m_K^0)
\frac{E_{\bar d} dE_{\bar d}}{E_u} \cong
$$
$$
\cong 2 \frac{\bar {\mid M \mid^2}}{4 \pi} \frac{E_d}{m^2_{K^0}} ,
\eqno(18)
$$
where $E_d$ is given by expression (17).
\par
Then using the expression (16) for $\bar \mid M \mid^2$, one can rewrite
equation (4) in the form:
$$
W(K^0 \leftrightarrow \bar K^0) \cong
\frac{G^2 f^2_{K^0} (m_s + m_{\bar d})^2 m_{K^0}}{8\pi}  ,
\eqno(19)
$$
and
$$
\tau_0 = \frac{1}{W_0(... )} .
$$
\par
Then the time $\tau(...)$ of $K^0  \leftrightarrow \bar K^0$
transition is
$$
\tau(K^0  \leftrightarrow \bar K^0) =
\frac{1}{W(K^0  \leftrightarrow \bar K^0)} .
\eqno(20)
$$
The mass difference of $K^0_1 - K^0_2$ is
$$
\Delta m = m_1 - m_2 = \cong \frac{h}{\tau_{K^0}}  .
\eqno(21)
$$
The length $L$ of  $K^0  \leftrightarrow \bar K^0$ oscillations is
$$
L_{phas} = \tau_{K^0}(...) v_{K^0} ,
\eqno(22)
$$
where $v_{K^0}$ is the velocity of $K^0$ meson.
\par
\noindent
and
$$
L_{phas} = \tau V = \frac{8 \pi p_{K^o}}{G^2 f^2_{K^o} m^2_s m^2_{K^o}} .
\eqno(23)
$$
Transition to the length of $K^o$ oscillations for $L_{phys}$ in
the model of dynamical analogy of Cabibbo-Kobayashi- Maskawa matrices
is given by Exp. (22) by using Exp. (6). \\

{\bf C) The scheme of $K^0 , \bar K^0$ oscillations}\\

\par
As an example of this oscillation we consider the oscillation of $K^0$
mesons produced in the reaction $\pi^- + P \to K^0 + \Lambda$.
At $t = 0$ there is the state $K^0 (0)$, then in time $t \ne 0$ for $K^0 (t)$, if
to take into account equation (10), we get $K^0$:
$$
K^0(t) = \frac{1}{2} [ ( K^0 + \bar K^0) exp(-im_1 t - \frac{\Gamma_1 t}{2})
 + (K^0 - \bar K^0) exp(-im_2 t - \frac{\Gamma_2 t}{2})]  =
$$
$$
= \frac{1}{2} K^0 exp(-i m_2 t)
[exp(-i \Delta m t -\frac{\Gamma_1}{2}t) +
 exp(- \frac{\Gamma_2}{2}t)] +
\eqno(24)
$$
$$
+ \frac{1}{2} \bar K^0 exp(-i m_1 t)
[exp(i \Delta m t -\frac{\Gamma_1}{2}t) + exp(- \frac{\Gamma_2}{2}t)] .
$$
From (23) it is clear that on the background of $K^0$ meson decays,
the oscillations of $K^0$ mesons take place [8, 10].

\section{ Comparisons of  $K^0  \leftrightarrow \bar K^0$ Oscillations in
the Diagram and the Phase Volume Approaches}

The length of $K^o \leftrightarrow \bar K^o $  in diagram approach
is [11]
$$
L_{diag} = 2 \pi \frac{2 p_{K^o}}{| m^2_1 - m^2_2 |} =
\pi \frac{2 p_{K^o}}{m_{K^o} \Delta m} ,
\eqno(25)
$$
where $\Delta m$ is
$$
\Delta m = \frac{8}{3} m_{K^o} f^2_{K^o} G  ,
\eqno(26)
$$
and $G$ is given by expression (12).
\par
The relation of these two lengths is:
$$
\frac{L_{osc}}{L_{phas}} = \frac{3}{32} G m^2_s      ,
\eqno(27)
$$
and
$$
\frac{L_{osc}}{L_{phas}} << 1     ,
\eqno(28)
$$
i.e., the length of $K^0  \leftrightarrow \bar K^0$
oscillations in the phase volume approach  in the standard
model and in the model of dynamical analogy of
Cabibbo-Kobayashi- Maskawa matrices is much more
than the length of these oscillations in the diagram approach and
in the model of dynamical analogy of Cabibbo-Kobayashi- Maskawa matrices.
It means that on the background of $K^o$ decays it is hard to register
the phase volume $K^o$ oscillations in contrast to $K^o$ oscillations
in the diagram approach.

\section{Conclusion}

The elements of the theory of dynamical expansion of the theory of weak
interaction working at the tree level, i.e. the model of dynamical analogy
of Cabibbo-Kobayashi- Maskawa matrices, were given.
\par
The equation for mass difference of $K^0_1, K^0_2$ mesons or the length
of $K^0, \bar K^0$ meson oscillations in the phase volume approach in
the frame work of the standard model of the weak interactions and
the model of dynamical analogy of Cabibbo-Kobayashi- Maskawa matrices
have been calculated (in this model the oscillations of $K^0, \bar K^0$
mesons arise at violation of strangeness by $B$ bosons).
Comparison of the length of oscillations in the diagram and phase volume
approaches was done. The length of $K^0  \leftrightarrow \bar K^0$
oscillations in the phase volume approach  is much more
than the length of these oscillations in the diagram approach.\\

\par
{\bf References}  \\

\par
\noindent
1. Beshtoev Kh.M., JINR, E2-94-293, Dubna, 1994;
\par
Turkish Journ. of Physics 1996, 20, p.1245;
\par
JINR, E2-95-535, Dubna, 1995;
\par
JINR, P2-96-450, Dubna, 1996.
\par
JINR Commun. E2-97-210, Dubna, 1997.
\par
\noindent
2. Review of Particle Prop., Phys. Rev. 1992, D45, N. 11.
\par
\noindent
3. Cabibbo N., Phys. Rev. Lett., 1963, 10, p.531.
\par
\noindent
4. Kobayashi M. and Maskawa K., Prog. Theor. Phys., 1973, 49,
\par
p.652.
\par
\noindent
5. Maiani L., Proc.Int. Symp.  on  Lepton-Photon Inter., Hamburg,
\par
   DESY, p.867.
\par
\noindent
6. Glashow S., Iliopoulos J. and Maiani L., Phys. Rev., 1970,
\par
D2, p.1285.
\par
\noindent
7. Beshtoev Kh.M., JINR Commun. E2-93-167, Dubna, 1993.
\par
Chinese Journal of Phys. 1996, v.34, p.979.
\par
\noindent
8. Okun L.B., Leptons and Quarks, M., Nauka, 1990.
\par
\noindent
9. Buras A.J., Harlander M.K., MPI--PAE/TTh/92, TUM--T31--25/92.
\par
\noindent
10. Beshtoev Kh.M., Fiz. Elm. Chastitz At. Yadra, 1996, v.27, p.23.
\par
\noindent
11. Beshtoev Kh. M., JINR Commun. E2-98-125, Dubna, 1998.

\end{document}